\documentclass[12pt]{article}
\renewenvironment{thebibliography}[1]
 { \rm
   \begin{list}{\arabic{enumi}.}
    {\usecounter{enumi} \setlength{\parsep}{0pt}
     \setlength{\itemsep}{3pt} \settowidth{\labelwidth}{#1.}
     \sloppy
    }}{\end{list}}

\def\A{{\bf A}}  

\def\x{{\bf x}}

\def\y{{\bf y}}
\def\k{{\bf k}}

\def\E{{\bf E}}

\newcommand{\gapproxeq}{\lower.7ex\hbox{$\;\stackrel{\textstyle>}{\sim}\;$}}
\newcommand{\lapproxeq}{\lower.7ex\hbox{$\;\stackrel{\textstyle<}{\sim}\;$}}

\usepackage{graphicx}
\usepackage{dcolumn}
\usepackage{bm}

\def\lsim{\mathrel{\rlap{\lower4pt\hbox{\hskip1pt$\sim$}}
    \raise1pt\hbox{$<$}}}
\def\gsim{\mathrel{\rlap{\lower4pt\hbox{\hskip1pt$\sim$}}
    \raise1pt\hbox{$>$}}}

\begin{document}
\titlepage

\vglue 1cm

\begin{center}
{{\bf THE LOW LYING GLUEBALL SPECTUM } 

\vglue 1.0cm
{Adam P. Szczepaniak\\}

\bigskip
{\it Department of Physics and Nuclear Theory Center \\
 Indiana University, Bloomington, IN 47405, USA \\}}

\vglue 1.0cm
{Eric S. Swanson\\}

\bigskip
{\it Department of Physics and Astronomy, University of Pittsburgh, \\
Pittsburgh PA 15260,  \\
Jefferson Lab, 12000 Jefferson Ave,
Newport News, VA 23606 \\}

\vglue 0.8cm

\end{center}

{\rightskip=2pc\leftskip=2pc\noindent
}
 
\vskip 1 cm

The complete low-lying positive charge conjugation glueball
spectrum is obtained from QCD. The formalism relies on the construction
of an efficient quasiparticle gluon basis for  Hamiltonian QCD in Coulomb
gauge. 
The resulting rapidly convergent Fock space expansion is exploited to 
derive quenched low-lying glueball masses with no free parameters which are
in remarkable agreement with lattice gauge theory.

\newpage 
\baselineskip=20pt

{\it Introduction } The scalar glueball has been called the fundamental particle of QCD\cite{geoff}. Indeed, 
its existence and nonzero mass are a direct consequence of the
nonabelian nature of QCD and the confinement phenomenon. It is clear that
finding and understanding the scalar (and other) glueballs is a vital step
in mastering low energy QCD.

Recently quenched lattice gauge theory has been able to 
 determine the low lying 
glueball spectrum with reasonable accuracy~\cite{MP} (only very preliminary
determinations of other matrix elements have been attempted). 
 These data serve as a useful benchmark in the development of a qualitative
model of the emergent properties of low energy QCD.
The models may then be used to guide experimental glueball searches.

Previous models of glueballs have relied on {\it ad hoc} effective
degrees of freedom such as flux tubes~\cite{IP}, bags~\cite{bm}, or constituent
gluons~\cite{gbs}.
We note that some of the models listed in Ref.~\cite{gbs} construct states with
massive gluons, while others either use transverse gluons or dynamically
generated gluons masses. Models in the former category contain spurious states due
to the presence of unphysical longitudinal gluon modes.
Sum rule computations of glueball properties exist~\cite{HJZ},
however, they are based on phenomenological properties of the
spectrum. 
Finally, the conjectured duality between supergravity and large-N gauge 
theories has been used to compute the glueball spectrum in 
non-supersymmetric Yang-Mills theory by solving the supergravity wave equations in a black hole geometry~\cite{ads}.
Unfortunately all of these approaches suffer from weak or conjectured connections
to QCD. 

We present a computation of the positive charge conjugation
glueball spectrum which  arises from QCD and is 
 systematically improvable.  
The computation is based on the formalism presented in Ref.~\cite{ss7} 
 in which
the QCD Hamiltonian in Coulomb gauge is employed as a starting point. 
 Coulomb gauge
 is efficacious for the study of bound states because all degrees of 
 freedom are
physical (there are no ghost fields in this gauge) and a positive definite norm
exists~\cite{cg}. Furthermore, resolving the Coulomb gauge constraint 
 produces an instantaneous
interaction (the nonabelian analogue of the Coulomb interaction)
which, as shown in Ref.~\cite{ss7} 
 may  be used to generate  bound states. Because the temporal component of the vector potential is renormalization
group invariant in Coulomb  gauge (this is not true in other gauges), the instantaneous
potential does not depend on the ultraviolet regulator or the renormalization
scale~\cite{Z}. This fact permits a physical interpretation of the 
 instantaneous potential which is a central aspect of our formalism.

The pure gauge QCD Hamiltonian may be written as~\cite{cg} 
 $H_{QCD} = H_0 + \delta H$ with

\begin{equation}
H_0 = {1 \over 2}\int d\x  \left[ {\bm E}^2 + {\bm B}^2 \right ]
+ {1\over 2}\int d\x d\y\, \rho^a(\x) K^{(0)}(\x-\y) \rho^a(\y)
\label{h0}
\end{equation}
and
\begin{equation}
\delta H = V_{3g} + V_{4g} + V_J + V_C \label{hint}
\end{equation}
\noindent
Here ${\bm B} = {\bf \nabla} \times {\bf A}$ is the abelian part of
the chromomagnetic field and ${\bm E} = -\partial/\partial t \A$ is the
 chromoelectric field. The third term in $H_0$ represents the
 nonabelian, instantaneous Coulomb interaction between color charges, 
 $\rho^a = -f^{abc}\E^b\cdot\A^c$, mediated by an effective potential $K^0$ 
 computed by taking a vacuum expectation value of the Coulomb kernel,
\begin{equation}
K^0(\x - \y)\delta_{ab} = g^2 \langle \Psi| 
\left[ ({\bm \nabla}\cdot {\bm D})^{-1} (- \nabla^2)
({\bm \nabla}\cdot {\bm D})^{-1}  \right]_{\x,a;\y,b} |\Psi \rangle. 
\end{equation}
with ${\bm D}^{ab}= \delta^{ab} {\bm \nabla} - g  f^{abc} \A^c$ being
the covariant derivative in the adjoint representation.   
For the vacuum wave functional,  $\Psi[{\bm A}] = \langle {\bm A} | \Psi
\rangle$ we take a variational ansatz,
\begin{equation}
  \Psi[{\bm A}] = \exp\left(-{1\over 2}\int {{d^3{\bf k}}\over {(2\pi)^3}} 
{\bf A}^a({\bf k}) \omega(k) {\bf A}^a(-{\bf k}) \right)
\end{equation}
with the variational parameter $\omega(k)$ determined by minimizing the
{\it vev} of $H$. The correction terms in $\delta H$ include  
$V_{3g}$ and $V_{4g}$ which are the three and four gluon operators originating
 from the difference between the full and the abelian chromomagnetic
 field.  $V_J$ denotes a contribution from the Faddeev-Popov
 determinant in the kinetic term. The effects of $V_J$ and the Faddeev-Popov
 determinant in the functional integrals have recently being studied
 in Ref.~\cite{FP} where it was found that its
 effects can be effectively included in the variational parameter
 $\omega(k)$. Finally, $V_C$ is the difference between the Coulomb
 operator and its {\it vev}, $K^0$. In the calculation of the glueball
 spectrum it results in operators mixing two and three, quasi-particle
 wave functions.  We note that after renormalization the coupling $g$ 
 is absorbed into the Faddeev-Popov operator, which then defines the
 Coulomb gauge analog of a ghost propagator~\cite{ss7, FP}. 
 The renormalized effective potential $K^0$ is fixed  by comparing
 with the quenched lattice QCD static potential. 
A very accurate representation
 of the static confinement potential is achieved~\cite{ss7}.

 The variational vacuum defined above also specifies a Fock
 space of quasiparticle excitations corresponding to effective 
 gluonic degrees of freedom (which we call quasigluons). 
 Such quasigluons obey ``massive'' dispersion relation in the
 variational vacuum and therefore improve the description of gluonic 
 bound states since mixing between states with
 different number of quasiparticles is suppressed due to their effective
 mass.  


The calculation of the ${\it vev}$ of the Hamiltonian and the
properties of the quasiparticle excitations were discussed in
Refs.~\cite{ss7, ss6, FP}. These require solving a set of coupled integral
 Dyson equations and as a result one finds that the function $\omega(k)$, 
 which in a free theory is given by $\omega(k) = k$, becomes finite as 
 $k\to 0$. The value $\omega(0)$ can be related to the slope of the
 static potential at large distances.


{\it Fock Space Expansion and the Glueball Spectrum.}  
The quasigluons which emerge in the analysis of Ref.~\cite{ss7} set
the QCD scale parameter via the low momentum dispersion relation $ r_0 \omega(k \to 0) = 1.4$, where $r_0$ is
the lattice Sommer parameter. Using the Regge string tension or $\rho$ 
 mass to fix the scale then
gives $\omega(0) = 600-650$ MeV.  It is natural to interpret this scale as a
dynamical gluon mass~\cite{jc}. Thus the formalism of Ref. \cite{ss7} provides
a justification of a Fock space expansion in terms of quasigluons and
gives the leading instantaneous interaction between the quasigluons. 
 In view of this it is natural to attempt a description of low lying 
 glueballs in the pure gauge sector of QCD.

In this approach positive charge conjugation glueballs are dominated by the two
quasigluon contribution. These may mix with three and higher
quasigluon states via transverse gluon exchange (and, in general, via
any term in $\delta H$). Mixing with single quasigluon states is
excluded because color nonsinglet states are removed from the spectrum
due to infrared divergences in the color non-singlet spectrum~\cite{ss6}. 
Finally, the scalar glueball is orthogonal to 
the vacuum due to the form of the gap equation.


The resulting bound state equations are shown in Eq.~(\ref{BSEqn}).
There is one orbital component of glueball wave function for 
$J^P=0^+$ and two for  $J^P=({\rm even} \ge 2)^+$. 
 These are denoted by $\psi_i(k), i=1,2$. 
The first term on the right hand side of Eq.~(\ref{BSEqn}) represents the 
quasigluon kinetic energy (the gluon gap equation has been employed
to simplify the expression),
the second term is the quasigluon self energy, and third
represents the interaction between quasigluons in the channel of interest.

\begin{eqnarray}
& & \int {{k^2 dk} \over {(2\pi)^3}} 2 \omega(k) |\psi_i(k)|^2
  + {N_C \over 2} \sum_i  \int {{k^2 dk}\over {(2\pi)^3 }} 
 {{q^2 dq}\over {(2\pi)^3 }} {{\omega(k)}\over {\omega(q)}} 
\left[ {4\over 3} V_0 + {2\over 3} V_2 \right] |\psi_i(k)|^2
\nonumber \\
& &  - {N_C \over 4 } \int {{k^2 dk}\over {(2\pi)^3 }} 
 {{q^2 dq}\over {(2\pi)^3 }} 
{\left(\omega(k) + \omega(q)\right)^2\over \omega(k) \omega(q)}
 \psi^{*}_i(q) K_{ij}(q,k) \psi_j(k)  
 = E \int {{k^2 dk} \over {(2\pi)^3}}|\psi_i(k)|^2 \nonumber \\y
\label{BSEqn}
\end{eqnarray}


with 
\begin{eqnarray}
K_{11} &=& {3J^2 + 3J -2 \over (2J-1)(2J+3)} V_J
+ { {J(J-1)}\over {2(2J-1)(2J+1)}} V_{J-2} \nonumber \\
&+& { {(J+1)(J+2)}\over {2(2J+3)(2J+1)}} V_{J+2} 
\end{eqnarray}

\begin{eqnarray}
K_{22} &=& { 3(J+2)(J-1) \over (2J-1)(2J+3)} V_J +
{{ (J+2)(J+1)}\over {2(2J+1)(2J-1)}} V_{J-2} \nonumber \\
&+& {{ J(J-1)}\over {2(2J+1)(2J+3)}} V_{J+2} 
\end{eqnarray}

and 

\begin{eqnarray}
K_{12} &=& K_{21} = \sqrt{ (J-1)J(J+1)(J+2) }  \nonumber \\
&\times& \Big[ {1 \over {2(2J+3)(2J+1)}} V_{J+2}
+ {1\over {2(2J+1)(2J-1)}} V_{J-2}  \nonumber \\
&-& { 1\over {(2J+3)(2J-1)}} V_J \Big]  
\end{eqnarray}

The bound state equations for other glueballs are as in
Eq. \ref{BSEqn}, with the exception that the wave function
index takes on a single value. For these cases the interaction kernels are
as follows:

\noindent $J^P = (odd\ge 3)^+$ (there is no $1^+$ gg glueball)  
\begin{equation}
K = {{J+2}\over {2J+1}} V_{J-1} + {{J-1} \over {2J+1}} V_{J+1}
\end{equation}

\noindent $J^P = (even\ge 0)^-$ 
\begin{equation}
K = {{J}\over {2J+1}} V_{J-1} + {{J+1} \over {2J+1}} V_{J+1}
\end{equation}

\noindent $J^P = 0^+$ 
\begin{equation}
K = {2\over 3}\left( V_0 + {V_2\over 2} \right)
\end{equation}

In all these relations the interaction is defined as

\begin{equation}
V_L(q,k) = 2\pi \int_{-1}^{1} dx K^{(0)}(q,k,x={\hat{\bf k}}\cdot {\hat{\bf q}})P_L(x) 
\end{equation}

\noindent
and the potential $K^{(0)}(q,k,x)$ is that derived in Ref.~\cite{ss7} 
 with the QCD scale chosen to be $\omega(0) = 600$ MeV.
Finally we note that there are no $J^P = (odd)^-$, or $C=-$ glueballs
 at lowest order in the Fock space expansion.

{\it Higher Order Corrections.}  It is of course desirable to test 
the efficacy of the Fock space expansion
employed here by explicitly checking the effect of coupling to the
three or higher  quasigluon spectrum. This is a difficult coupled
channel problem and we therefore content ourselves  with a
perturbative evaluation of these effects in this initial study. 
 Specifically, the energy shift $\delta E_n = \sum_m |\langle gg|
 \delta H |m(ggg)\rangle |^2/(E_n-E_m)$ must be evaluated. 
 Duality implies that when the 
 energy transfer is  large, $(E_n-E_m) > \Lambda$, where $\Lambda$ 
 is of the order of the QCD scale,  this sum may be evaluated in its 
 perturbative form (with partonic gluons in the intermediate state). 
 We compute here the effects of the three-gluon coupling from $\delta
 H$. This is the leading interaction in terms of expansion in the
 coupling constant. After renormalization $g^2/4\pi \to \alpha(p^2)$ where 
$p^2$ represents a characteristic momentum in integrals when
 computing matrix elements. The running coupling expansion for the
 remaining sum over low energy modes is certainly less justifiable,
 however as shown in examples in Ref.~\cite{ss7}, such soft
 corrections also seem to be small. For numerical efficiency the low
 energy part of the guasigluon exchange was approximated with a local 
 four-gluon interaction (we note that this effective interaction also 
 accounts for the four-gluon interaction present  in the Hamiltonian)

\begin{eqnarray}
V_c =  &&C(\Lambda) f^{abc}f^{ade} \int {d^3k_1\over (2 \pi)^3}
{d^3k_2\over (2 \pi)^3} {d^3k_3\over (2 \pi)^3} {d^3k_4\over (2
  \pi)^3} (2\pi)^3 \delta(\k_1+\k_2+\k_3+\k_4) \cdot\nonumber\\
&& \exp\left(-(k_1^2+k_2^2+k_3^2+k_4^2)/\Lambda^2\right) 
 A_i^b(\k_1) A_j^c(\k_2) A_i^d(\k_3) A_j^e(\k_4).  \nonumber \\
\label{local}
\end{eqnarray}

\noindent
where $C$ is a dimensionless parameter of the order of $g^2(\Lambda)
\sim 1$. Standard effective field theory techniques were subsequently 
employed: the factorization scale $\Lambda$ was chosen and the 
 coupling $C$ was fixed by comparison to the lattice scalar glueball
 mass. Other mass predictions then follow. The scale $\Lambda$ was
 then varied to ensure that the procedure is stable and that the
 coupling remains `natural' (of order unity).  To maintain consistency 
 the effect of these terms should be incorporated into the quasigluon
 gap  equation and the gluon self energy. However, it may be shown that
the effect of contact terms are canceled in the bound state equation when 
 the gap equation is used to simplify the quasigluon kinetic and self energies.
 This is not true for the high-momentum gluon exchange terms which add
 a UV dominated correction to the single gluon kinetic energy. These
 have negligible effect on low energy spectrum and have been ignored.

\begin{center}
\begin{table}
\caption{Glueball Spectrum}
\begin{tabular}{|l|c|l|l|}
\hline\hline
state & this work & LGT & Ref \\
 & (no mixing) &  (GeV) & \\
\hline
$0^{++}$ & 1.98 &  1.73(5)(8) & \cite{MP}\footnote{The first error is combined statistical and systematic,the second is from scale fixing.} \\
 &  &  1.754(85)(86)  & \cite {LWC} \\
 &  &  1.627(83)  & \cite{NRW}  \\
 &  &  1.686(24)(10) &  \cite{GF11}  \\
 &  &  1.645(50)  & \cite{UKQCD}  \\
 &  &  1.61(7)(13)  & \cite{T} \\
$0^{++}{'}$ & 3.26 & 2.67(18)(13)  & \cite{MP} \\
$2^{++}$ & 2.42 &  2.40(2.5)(12) &  \cite{MP}\\
 &  &  2.417(56)(117) &  \cite{LWC}  \\
 &  &  2.354(95)  & \cite{NRW}\\
 &  &  2.26(12)(18)  & \cite{T}  \\
 &  &  2.380(67)(14)  & \cite{GF11}  \\
 &  &  2.337(100)  & \cite{UKQCD}  \\
$2^{++}{'}$ & 3.11 & 3.499(43)(35)  & \cite{MPo}   \\
$0^{-+}$ & 2.22 &   2.59(4)(13)  & \cite{MP}\\
         &      &   2.19(26)(18)  & \cite{T} \\
$0^{-+}{'}$ & 3.43 & 3.64(6)(18)  & \cite{MP}   \\
$2^{-+}$ & 3.09 &   3.10(3)(15)  & \cite {MP}   \\
$2^{-+}{'}$ & 4.13 & 3.89(4)(19)  & \cite{MP}   \\
$3^{++}$ & 3.33 &  3.69(4)(18)\footnote{Possible mixing with higher states.} & \cite{MP} \\
$3^{++}{'}$ & 4.29  &  &    \\
$4^{++}$ & 3.99 &  3.65(6)(18) & \cite{LW} \\
$4^{++}{'}$ & 4.28 &  &   \\
$4^{-+}$    & 4.27 &   &   \\
$4^{-+}{'}$ & 4.98 & &   \\
\hline\hline
\end{tabular}
\end{table}
\end{center}

{\it Results and Conclusions.} The 
lowest order predictions for the quenched positive charge
conjugation glueball spectrum are presented in Table~1. We stress that 
 there are no free parameters in this computation; the renormalization 
 group parameters and the scale were fixed by comparison to the Wilson
 loop static interaction~\cite{ss7}. Although one may anticipate
 splittings on the order of 100 MeV due to coupled channel effects,
 the general agreement with lattice data is quite good (the $\chi^2$
 per degree of freedom for the six measured lowest spin-parity states
 is 1.5). Nevertheless we note that the authors of Ref.~\cite{MP}
 state that the $3^{++}$ may have significant mixing with higher
 states and the quoted $4^{++}$ mass should be regarded as preliminary.

Although it appears to be difficult to push lattice mass computations 
above $4\mbox{ GeV}$  it would be interesting to measure the quenched 
$4^{-+}$ glueball mass to test the  prediction made in Table~1. Lastly
we note that all radial excitations lie roughly $1\mbox{ GeV}$ above 
 their respective ground states, except the $4^{++}$. We have no
 explanation for this curiosity, but note that it implies that
 lattice extractions of the $4^{++}$ mass must be made with great care.


We note that the degeneracy between parity states reported in the first of 
Refs.~\cite{gbs} is not seen here. We suspect that the degeneracy is due to 
the nonrelativistic expansion of the interaction kernel made in that reference.

As stated above, coupled channel effects are expected to modify the
spectrum at the $10-100 \mbox{ MeV}$ level.
As an initial estimate of the size of these effects we simply set $C = 0$ 
 (Eq.~\ref{local}) and evaluate the energy shift due to perturbative
 one gluon exchange. As expected, the scalar glueball mass experiences 
 the largest shift, with a reduction in mass of roughly $200\mbox{
   MeV}$  (from $1980$ to $1790\mbox{ MeV}$). It is gratifying that 
 this brings the scalar glueball into excellent agreement with lattice 
 gauge theory. We proceed by incorporating the effective contact interaction. 
 The factorization scale was varied 
 between $0.25$ and $10 \mbox{ GeV}$, stable results were found
 between  $1$ and $8 \mbox{ GeV}$, with a value of $C$ given roughly 
 by $-0.4$ in this range. We find that the tensor glueball mass is reduced
 by roughly $100 \mbox{MeV}$, while other masses experience somewhat 
smaller shifts.  Thus it appears that low lying glueballs are 
 indeed dominated by their two quasigluon Fock components.  However it
 is clear that a careful examination of coupled channel effects and
 better lattice data are required to make a definite  statement about
 the efficacy of our approach. 

Fluctuations of the topological charge density have pseudoscalar quantum
numbers. This raises the possibility that the QCD anomaly affects the lightest
$0^{-+}$ glueball mass. Topological effects have so far not been 
incorporated into our formalism. Doing so would require modification of the
vacuum ansatz to reflect the identification of gauge equivalent
field configurations at the boundary of the fundamental modular region.  This
allows contributions from field configurations with nonzero topological charge.
Indeed a cross-over between the $0^{-+}$ and $2^{++}$ glueball masses has
been observed on the lattice as a function of the renormalised
coupling~\cite{vanBaal} if boundary conditions are imposed.

Further aspects of the gluonic structure laid out in Ref. \cite{ss7} 
may be investigated by an examination of the adiabatic
potential surfaces of heavy quark hybrids (this probes nonperturbative 
gluon-confinement potential couplings). Extensions to the light 
hybrid spectrum will prove of interest to searches for these new
states at Jefferson Lab. Finally, the short range structure of the 
meson sector is dominated by coupled channel effects and
nonperturbative gluodynamics. The wealth of experimental information
in this sector will provide a definitive test of the dynamics being 
advocated in our approach.

{\it Aknowledgments} This work was supported by DOE under contracts
DE-FG02-00ER41135,  DE-AC05-84ER40150 (ES), and DE-FG02-87ER40365 (AS).

\vglue 0.4cm


\begin{thebibliography}{99}
\bibitem{jc}
The relationship of a dynamical gluon mass to vortex-driven confinement
and gauge symmetry breaking is thoroughly discussed by Cornwall; see,
for example, the second and third references of \cite{gbs}.
\end{thebibliography}
\end{document}